\documentstyle[sprocl,epsf]{article}
\begin{document}

\title{LEVEL STATISTICS IN THE QUANTUM HALL REGIME}
\author{M.~BATSCH$^{1,2}$ and L.~SCHWEITZER$^1$}
\address{$^1$Physikalisch-Technische Bundesanstalt, Bundesallee 100,\\ 
D-38116 Braunschweig, Germany\\
$^2$I. Institut f\"ur Theoretische Physik, Jungiusstra{\ss}e 9,\\ 
D-20355 Hamburg, Germany}
\date{\today}
\maketitle
\abstracts{The statistical properties of energy eigenvalues in the 
critical regime of the lowest Landau band are investigated. The nearest 
neighbour level spacing distribution $P(s)$ and the Mehta quantity $I_0$, 
from which the scaling exponent $\nu$ of the correlation length can be 
derived, are calculated. 
While $P(s)$ disagrees with recent analytical predictions, $\nu$ is 
found to be in agreement with theoretical and experimental results.} 

\section{Introduction}
In recent years the investigation of energy level statistics in systems 
showing a localisation-delocalisation transition has made remarkable progress. 
For the metallic side of the transition  the spectral 
fluctuations only depend on the symmetry of the system
(orthogonal, unitary or symplectic) and are well described by 
Random Matrix Theory (RMT)~\cite{Meh91}.  
On the insulating side, the energy levels are uncorrelated and exhibit 
Poissonian behaviour. 
At the transition point, new critical ensembles were 
found for the 3d orthogonal~\cite{AZKS88,Sea93,ZK95b} the 2d 
symplectic~\cite{SZ95,Eva95}, and the 3d unitary case~\cite{BSZK96}.
In the quantum Hall system no complete metal-insulator transition is 
observed, and the situation is unclear~\cite{HS92s,Ono95,FAB95,OO95}. 

This is due to the fact that the localisation length of the electronic 
wavefunctions diverges at the centre of the disorder broadened Landau bands. 
Hence, in the thermodynamical limit there is no range of extended 
states but only one energy where the states are multifractal
\cite{HS92s,PJ91,HKS92}. 
Therefore, the question is whether there exists a
scale-invariant level statistics in {\em finite systems} in the absence of 
true extended states. 

As a result of analytical theories~\cite{KLAA94,AKL94}
the asymptotic behaviour of the nearest neighbour level 
spacing distribution $P(s)$, where the energy difference $s$ 
is measured in units of the mean level spacing $\Delta $, was proposed to obey 
$P(s) \propto \exp(-c s^{2-\gamma})$ with
$\gamma = 1 - \frac{1}{\nu  d}$,
where $\nu $ is the critical exponent and $d$ the Euclidean dimension.
Fits to the numerical data in 3d orthogonal~\cite{ZK95b}, 
3d unitary~\cite{BSZK96} and 2d symplectic~\cite{SZ95,Eva95} systems did not 
result in a sufficient agreement. 
Investigations that seem to show a reasonable fit \cite{OO95,Eva94} 
require, however, values for $\nu$ which are different from those commonly 
accepted. Previously, a simple exponential decay of $P(s)$ was proposed 
\cite{AZKS88},
$P(s) \propto \exp(- \kappa  s)$, 
where $\kappa > 1 $ is a numerical constant. 
From the calculations in 3d orthogonal \cite{ZK95b} and 3d unitary systems 
\cite{BSZK96}, $\kappa \approx 1.9$, and for the 2d symplectic, 
$\kappa \approx 4.0$ was obtained \cite{SZ95}.
In our opinion, $\kappa$ should reflect the fractal structure
of the critical wavefunctions. 
Therefore, it is very interesting to study the asymptotics of $P(s)$ in 
the 2d QHE situation where no complete localisation-delocalisation transition 
occurs.
\begin{figure}[t]
\hspace*{1.cm}\epsfxsize8.75cm\epsfbox{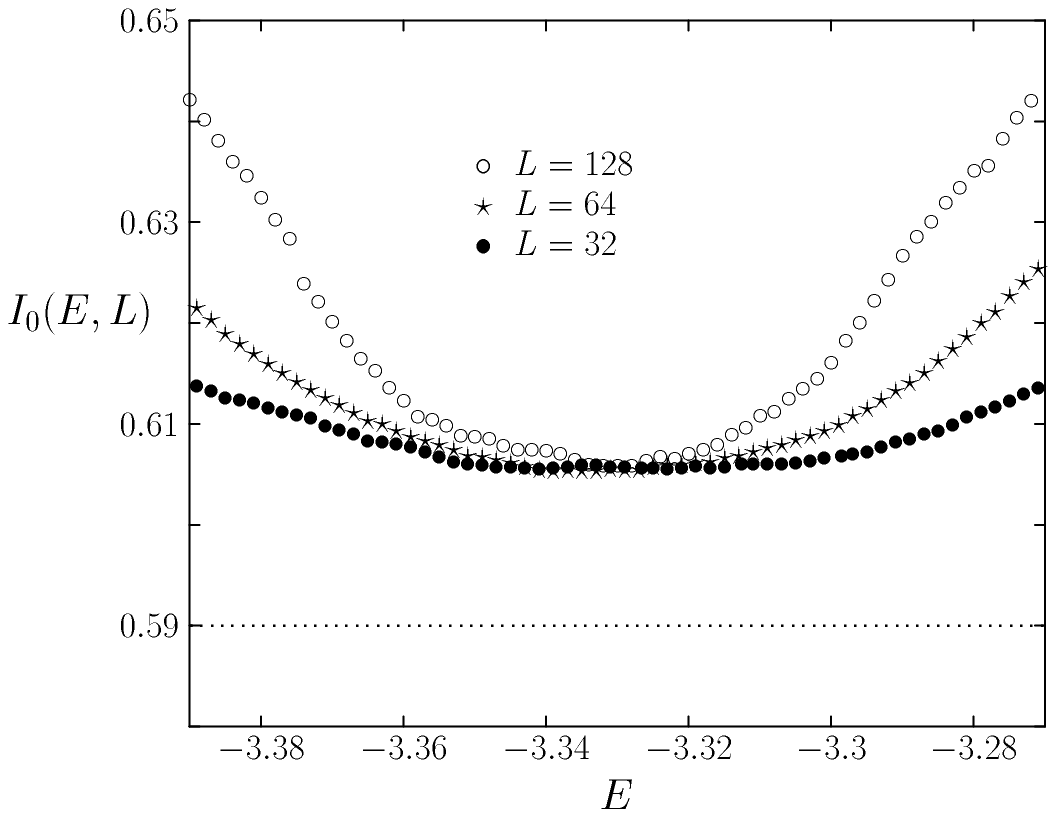} 
\caption[]{\label{I0}Energy and size dependence of $I_0$ at the critical 
energy in the lowest Landau band}
\end{figure}
\section{Model and method}
We investigate a 2d Anderson model with the magnetic field included via a 
phase-factor in the transfer matrix elements $V$ (see Ref.{\cite{HS92s}} 
for details). 
The diagonal disorder potentials are taken at random with constant 
probability from  $[ - W/2, W/2]$ and $W/V=2.5$.  
The strength of the magnetic field equals 1/8 flux quanta per unit cell. 
We diagonalise the corresponding Hamiltonian matrix for finite squares 
of size $L^2=
32^{2}$, $64^{2}$, $128^{2}$, and $160^{2}$ 
using a Lanczos procedure. To obtain a reliable statistic, up to 3$^6$
realizations were calculated. 
\section{Results and discussion}
To find the energy range of the eigenstates with localisation length
larger than the system size and to extract the critical exponent $\nu$, 
we employ the quantity $I_0$ known from RMT~\cite{Meh91}.
The calculation does neither require a fitting procedure as the Brody level 
repulsion parameter~\cite{Ono95}, nor some cutoff value, as for example the 
integrated $\Delta _{3}$ statistics~\cite{HS94}. $I_0$ is obtained by integrating over $P(s)$
\begin{equation}
I_{0} = \int_{0}^{\infty} \int_{s}^{\infty} \int_{s'}^{\infty} P(s'') ds'' 
ds' ds.
\end{equation}
The value of $I_{0}$ for the Gaussian unitary ensemble 
is $I_{0}=0.590$ and for uncorrelated (localised) levels $I_{0}=1$.
Multifractal states are expected to have a value larger than 0.59. From our 
calculations we find $I_{0}=0.605 \pm 0.002$ near the critical energy 
independent of the considered system sizes.
\begin{figure}[t]
\hspace*{1.cm}\epsfxsize8.75cm\epsfbox{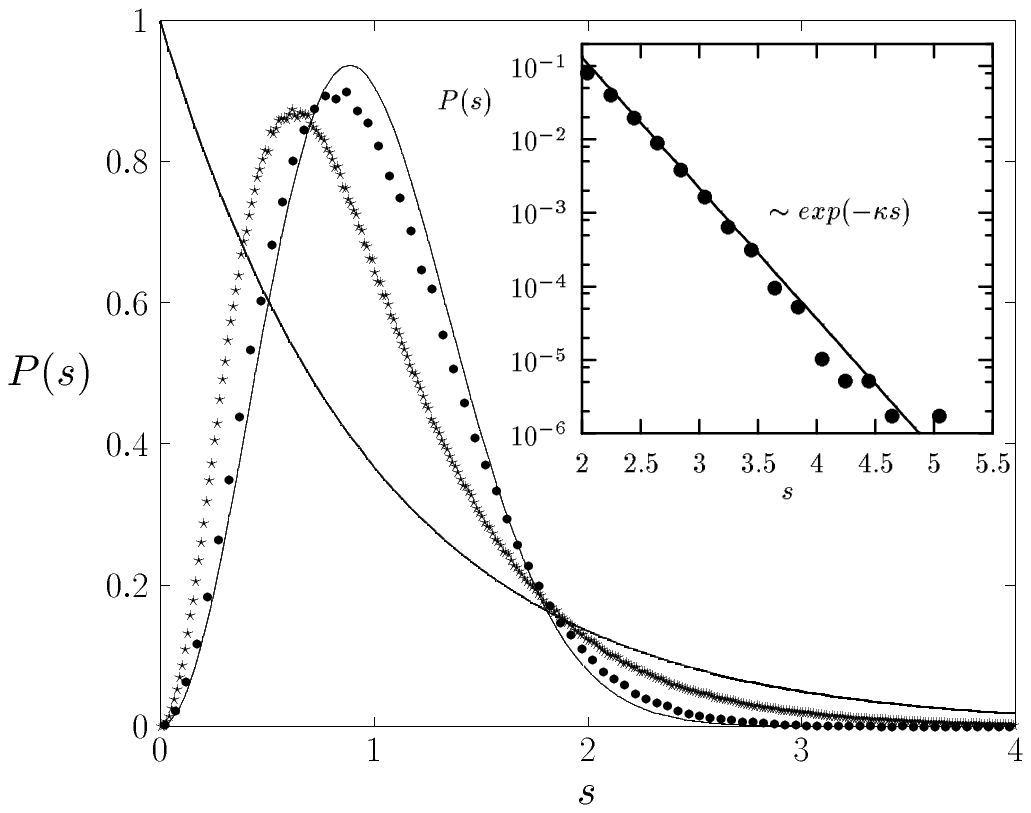} 
\caption[]{\label{pofs}The critical QHE-$P(s)$ ($\bullet$) in comparison with
the critical 3d unitary ($\star$), the Poisson and the GUE result. The inset 
shows a $\exp(-\kappa s)$ decay with $\kappa\approx4.1$}
\end{figure}
Fig.~\ref{I0} displays the energy dependence of $I_{0}$ in the energy range
about the critical point. We have carefully checked that $I_0$ does not 
dependent on the width of the energy interval from which the eigenvalues 
are taken.
For all system sizes one observes a minimum $I_0 \approx 0.605$ at the 
critical energy $E_{c}/V=-3.33$. 
States away from $E_{c}$ become more localised when the system size is 
enlarged.
From the scaling behaviour of $I_0(E,L)$ we extract the critical exponent
$\nu=2.4$ in good agreement with both theoretical 
\cite{HKS92} and experimental \cite{KHKP91} results.

The nearest neighbour level spacing distribution $P(s)$ is shown in 
Fig.~\ref{pofs}. It clearly deviates from the Wigner surmise although it 
shows a quadratic small-$s$ behaviour as expected for a unitary symmetry. 
Our result for the lattice model resembles that obtained for a continuum 
model \cite{Ono95,OO95}. Due to the large number of realizations we are able 
to extract the large-$s$ behaviour displayed in the inset of 
Fig.~\ref{pofs}. As for the other critical ensembles, it is not possible to
fit the tail of $P(s)$ by the proposal \cite{AKL94} 
$P(s)\sim\exp(-c s^{2-\gamma})$ 
with $\nu=2.4$, but it can be fitted by a simple exponential decay, 
$P(s) \propto \exp(-\kappa s)$, with $\kappa \approx 4.1$. 
This value is similar to the one found for a 
2d system with symplectic symmetry~\cite{SZ95}. 
In conclusion, there exists a critical level statistics in 2d QHE systems
despite the absence of 
a complete metal-insulator transition.

\section*{References}

\end{document}